\documentclass[a4paper,aps,pre,floatfix,twocolumn,nofootinbib,superscriptaddress,showpacs]{revtex4-1}

\usepackage[pdftex]{graphicx}
\usepackage{latexsym,amsmath}
\usepackage[pdftex]{color}
\usepackage{hyperref}
\usepackage{float}
\usepackage{microtype}

\newcommand{\av}[1]{\left\langle #1 \right\rangle}

\graphicspath{{/Users/romu/Dropbox/NoPAD/Voter/figures/}}

\begin{document}

\title{Generalized Voter-like model on activity driven networks with attractiveness}

\author{Antoine Moinet}

\affiliation{Aix Marseille Univ, Universit\'e de Toulon, CNRS, CPT,
  Marseille, France}

\affiliation{Departament de F\'isica, Universitat Polit\`ecnica de
  Catalunya, Campus Nord B4, 08034 Barcelona, Spain}

\author{Alain Barrat}

\affiliation{Aix Marseille Univ, Universit\'e de Toulon, CNRS, CPT,
  Marseille, France}

\affiliation{Data Science Laboratory, ISI Foundation, Torino, Italy}

\author{Romualdo Pastor-Satorras}

\affiliation{Departament de F\'isica, Universitat Polit\`ecnica de
  Catalunya, Campus Nord B4, 08034 Barcelona, Spain}

\begin{abstract}
  We study the behavior of a generalized consensus dynamics on a
  temporal network of interactions, the activity driven network with
  attractiveness. In this temporal network model, agents are endowed
  with an intrinsic activity $a$, ruling the rate at which they generate
  connections, and an intrinsic attractiveness $b$, modulating the rate
  at which they receive connections. The consensus dynamics considered
  is a mixed voter/Moran dynamics. Each agent, either in state $0$ or
  $1$, modifies his/her state when connecting with a peer. Thus, an active
  agent copies his/her state from the peer (with probability $p$) or
  imposes his/her state to him/her (with the complementary probability
  $1-p$). Applying a heterogeneous mean-field approach, we derive a
  differential equation for the average density of voters with activity
  $a$ and attractiveness $b$ in state $1$, that we use to evaluate the
  average time to reach consensus and the exit probability, defined as
  the probability that a single agent with activity $a$ and
  attractiveness $b$ eventually imposes his/her state to a pool of
  initially unanimous population in the opposite state. We study a
  number of particular cases, finding an excellent agreement with
  numerical simulations of the model. Interestingly, we observe a
  symmetry between voter and Moran dynamics in pure activity driven
  networks and their static integrated counterparts that exemplifies the
  strong differences that a time-varying network can impose on dynamical processes.
\end{abstract}


\maketitle

\section{Introduction}
\label{sec:intro}

A wide variety of complex physical systems are concerned with the
problem of an initially disordered configuration that is able to achieve
an ordered state by means of local pairwise dynamical
interactions. Examples of such systems range from the formation of an
opinion consensus in social systems~\cite{Castellano09} to the loss of
genetic diversity in evolutionary dynamics~\cite{drossel2001biological}.
These situations, implying a competition between different alternative
states diffusing among the agents, have been modelled with stochastic
copying or invasion processes.  In these models, each individual is
endowed with a state variable and copies or imposes his/her state from
or to neighbouring sites until one single state finally dominates the
whole system.  Among those different frameworks, the voter
model~\cite{liggett1985ips} was introduced to schematically model the
opinion spreading in human populations and has become emblematic for its
simplicity and analytical tractability. In this model, the agents
possess one of two discrete opinions, and at each time step an
individual is chosen and adopts the opinion of a randomly chosen
neighbour. On the other hand, in the context of evolutionary dynamics,
the Moran process~\cite{moran1962spe} considers a population of
individuals belonging to different species, that reproduce generating an
offspring that replaces a randomly chosen nearest neighbor. The voter
and Moran models differ thus in the direction in which the state is
transferred between pairs of interacting agents.

Recently, it has been acknowledged that the topology in which such
ordering dynamics takes place in real systems is often far for
homogeneous, and better represented in terms of a complex
network~\cite{barratbook,Newman2010}, in which agents are characterized
by a number of neighbors (degree) $k$ that is broadly distributed, with
a heterogeneous probability distribution $P(k)$ (degree distribution)
often schematically described by a power-law-like form
$P(k) \sim k^{-\gamma}$~\cite{Barabasi:1999}. This observation has led
to an intense research activity in order to unveil the different
properties of ordering dynamics in heterogeneous
topologies~\cite{Sood08,barratbook,PhysRevE.82.010103,1367-2630-10-6-063011,1742-5468-2009-10-P10024,2010arXiv1011.2395B,morettiheterovoter},
yielding a good understanding of the problem both at the numerical and
analytical levels.

These studies have mainly focused on the case of \textit{static}
networks, in which nodes representing agents are connected by a set of
edges, standing for pairwise possible interactions, that are fixed in
time and never change. However, many networks, and in particular social
ones, are dynamic in nature, given by a pattern of connections that
evolves in time. Such temporal networks~\cite{Holme:2011fk,Holme2015}
have been the subject of an intense research activity, considering in
particular their possible impact on the behavior of dynamical processes
running on top of
them~\cite{Holme2015,burstylambiotte2013,Isella:2011,PhysRevE.83.025102,min_spreading_2011,PhysRevE.85.056115}. Despite
their relevance, however, consensus dynamics have seldom been studied in
detail in temporal
topologies~\cite{Fernandez-Gracia2011,Karimi2013,Li:2018}.

Here we contribute to fill this gap by presenting a detailed study of
the voter and Moran processes on temporal networks, focusing on a
generalization of the recently introduced activity driven temporal
network model~\cite{2012arXiv1203.5351P}. In this
model~\cite{Alessandretti2017} agents are assigned an activity parameter
$a$ that determines their propensity to establish social interactions
with other individuals, and another variable $b$ defining their
attractiveness, which in turn determines the probability that they are
chosen by an active agent to interact.  We provide a full analysis of
basic ordering dynamics through a heterogeneous mean-field approach that
allows us to describe the dynamics of the process in the limit of a
large system size, and in particular to compute the average time to
reach consensus starting from a random configuration. Another quantity
of interest when studying consensus or invasion processes is the
so-called exit probability, defined as the probability that a single
agent having a discrepant opinion among an unanimous population manages
to spread his/her opinion to the whole population. In our work this
quantity plays a significant role as it highlights an interesting
symmetry between the voter model and the Moran process when comparing
the unfolding of these processes either on static heterogeneous networks
or on and the activity driven network. In particular, we show how,
depending on the type of dynamics, the effect of a node's
characteristics on the dynamics can be similar or drastically different
when the dynamics runs on a temporal network or on the corresponding
static aggregated network.

The paper is organized as follows: In Sec.~\ref{sec:voter-models-temp}
we define the variation of the voter model we consider, and the class of
activity driven network with attractiveness we use as a substrate for
the dynamics. In Sec.~\ref{sec:poiss-activ-driv} we provide a full
description of the dynamics and in particular we exhibit analytical
expressions for the exit probability and the average consensus time. In
Sec.~\ref{sec:part.cases} we give more insights for a few particular
forms of the joint distribution of the activity and attractiveness
$\eta(a,b)$.  Sec.~\ref{sec:numerics} presents numerical simulations
showing agreement with our analytical expressions, which are further
studied in some asymptotic limit of the activity distribution in
Sec.~\ref{sec:asympt}. Finally we conclude our work discussing our
results and exploring perspectives in Sec.\ref{sec:conclusions}.

\section{Consensus dynamics in generalized activity driven networks}
\label{sec:voter-models-temp}

\subsection{Activity driven networks with attractiveness}

We focus on the class of activity driven temporal networks
\cite{2012arXiv1203.5351P,starnini_topological_2013,Alessandretti2017},
which are based in the key ingredient that the formation of social
interactions is driven by some innate \textit{activity} of individuals,
which determines their tendency to establish interactions and is
empirically observed to be heterogeneously distributed
\cite{2012arXiv1203.5351P}. In this paradigmatic model, a fixed
population of $N$ agents is considered, each endowed with an
\textit{activity} $a$, representing the rate (probability per unit time)
at which s/he becomes active and draws edges (interactions) towards
other agents. In the original formulation of the
model~\cite{2012arXiv1203.5351P}, the chosen agents are selected
uniformly at random among all peers. In order to make the model more
realistic, one can assign to each agent a parameter $b$, called
\textit{attractiveness}, such that his/her probability of being selected
by an active peer is proportional to $b$ \cite{Alessandretti2017}.  The
activity $a$ and attractiveness $b$ are extracted at random for each
agent from a given joint distribution $\eta(a,b)$.

\subsection{Consensus dynamics}
\label{sec:voter-dynam-activ}

Coupling a dynamical process with a temporal network always entails the
problem of how to deal with the different time scales inherent in the
process and in the evolution of the network. Here we consider the
simplest case of a single time scale, imposed by the network
evolution. In this way, the state of an agent can only change when s/he
interacts with another agent, and is constant in the latency times
between interactions. The consensus dynamics are thus defined as
follows:
\begin{itemize}
\item We start from an initial configuration of states
  $s_i \in \{ 0, 1\}$, assigned to each agent.
\item In an interval of time $\delta t$, an agent $i$, in state $s_i$,
  becomes active with probability $a_i\delta t$, and chooses as peer
  another agent $j$ (in state $s_j$) with probability
  $\dfrac{b_j}{\sum_\ell b_\ell}$
\item The states $s_i$ and $s_j$ are updated according to the chosen
  consensus dynamics.
\item Time is updated $t \to t+\delta t$.
\end{itemize}

We consider three different variations of social dynamics, based on the
update dynamics of the state variables. Assuming that, at given time
$t$, agent $i$ becomes active, and chooses agent $j$ to start an
interaction, we consider the three different updates:
\begin{enumerate}
\item Voter dynamics: $ s_i := s_j\nonumber$ (i.e., $i$ adopts $j$'s state).
\item Moran dynamics: $s_j := s_i$.
\item Mixed dynamics: With probability $p$, $s_i := s_j$; with the
  complementary probability $1-p$, $ s_j := s_i$.
\end{enumerate}
In what follows, we consider the mixed update rule,
as the voter model and the Moran process are particular cases of the
latter obtained by setting $p=1$ or $p=0$ respectively.

\section{Heterogeneous mean-field analysis}
\label{sec:poiss-activ-driv}

When agent activation is ruled by a Poisson process, it is possible to
tackle the behavior of voter-like dynamics by extending the
heterogeneous mean-field \cite{barratbook,dorogovtsev07:_critic_phenom}
approach developed in Refs.~\cite{PhysRevLett.94.178701,Sood08} to study
voter dynamics on static networks. This method is based in a
coarse-graining of the network, considering that the state of an agent
with activity $a$ and attractiveness $b$ depends exclusively on those
two quantities. In this way, one considers a fundamental description in
terms of the fraction $\rho_{a,b}(t)$ of agents with activity strength
$a$ and attractiveness $b$ in the state $1$ at time $t$; in other words,
$\rho_{a,b}(t)$ is the probability that a randomly chosen agent with
activity $a$ and attractiveness $b$ is in state $1$ at time $t$. The
corresponding fraction of agents in state $0$ is given by the
complementary probability $1 - \rho_{a,b}(t)$.  The total fraction of
agents in state $1$, $\rho(t)$, is given by
\begin{equation}
  \rho(t) = \sum_{a,b} \eta(a,b) \rho_{a,b}(t).
  \label{eq:rho}
\end{equation}
To alleviate notation, we denote the pair $(a,b)$ by the symbol $h$,
writing thus $\rho_{a,b}(t) \equiv \rho_{h}(t)$.

The relevant functions defining the dynamics are the transition
probabilities $R_h$ and $L_h$ for respectively increasing and decreasing the number
of voters in state $1$, among the pool of agents with activity strength $a$
and attractiveness $b$, in a time interval $\delta t$. From these
transition probabilities, a differential equation ruling the evolution
of $\rho_h(t)$ can be derived, as well as information about the exit
probability and the average ordering time. In the dynamical rules
described in the previous section, agents activate independently so that
a priori multiple activations are possible during a single time
step. However the use of the transition probabilities $R_h$ and $L_h$
relies on the implicit hypothesis that only one flip attempt may occur
during a single time step, thus in order to ensure the validity of our
analysis, we impose that $\av{a}N\delta t\ll 1$ so that the probability
of counting more than one activation during $\delta t$ is almost zero. This is of course
always possible as the time step $\delta t$ is arbitrary.

Let us now derive the time evolution equation of the fraction
$\rho_{h}(t)$ of agents with activity strength $a$ and attractiveness
$b$ in state $1$ at time $t$.

\subsection{Evolution equation}
\label{sec:voter-model}

In a single time step, the number of agents with activity strength $a$
and attractiveness $b$ in state $1$ may either increase by one unit with
probability $R_h$, decrease by one unit with probability $L_h$, or stay
unchanged with probability $1-R_h-L_h$. Thus on average the variation
$\delta \rho_h$ reads
\begin{equation}
 \delta \rho_h = (+1)\times \dfrac{R_h}{N_h}+ (-1)\times
 \dfrac{L_h}{N_h} + 0\times\dfrac{1-R_h-L_h}{N_h} ,
\end{equation}
where $N_h$ is the number of agents in the state $(a,b)$.  In the
continuous time limit (for $\delta t \ll 1$) we may write
\begin{equation}
  \frac{\partial\rho_h(t)}{\partial t}=\dfrac{R_h-L_h}{N_h\,\delta t}.
\end{equation}
We consider the mixed process in which every agent, when activated,
might either copy the state of his/her peer with probability $p$, or impose
his/her own state to him with probability $1-p$. The transition
probabilities are thus given by 
\begin{eqnarray}
R_h &=& p\,R_{h}^{V}+(1-p)R_{h}^{M}\\
L_h &=& p\,L_{h}^{V}+(1-p)L_{h}^{V},
\end{eqnarray}
where the rates $L_h^X$ and $R_h^X$ refer to the voter ($X=V$) and Moran
($X=M$) dynamics, respectively.

In the case of the voter dynamics, these transition probabilities take the
form
\begin{eqnarray}
  R_h^{V} &=& N_h\,a\,\delta
              t\,(1-\rho_h)\dfrac{\av{b\,\rho_h}}{\av{b}},\label{eq:1}\\ 
  L_h^{V} &=& N_h\,a\,\delta
              t\,\rho_h\left(1-\dfrac{\av{b\,\rho_h}}{\av{b}}\right).\label{eq:2} 
\end{eqnarray} 
The origin of these expressions is easy to see. For example, in
Eq.~\eqref{eq:1}, the probability that the number of agents in state
$1$, activity $a$ and attractiveness $b$ increases by one unit is
proportional to the number of agents in this class in state $0$,
$N_h[1-\rho_h(t)]$, times the probability that any one of them becomes
active in a time interval $\delta t$ ($a \delta t$), times the probability that an
active agent generates a link to an agent  in state $1$, thus copying the
state of this last agent. The latter is the sum over all the agents $i$
of the probability that $i$ is chosen and is in state $1$, i.e.,
$\sum_i \frac{b_i}{\av{b}N}\,s_i = \frac{\av{b\,\rho_h}}{\av{b}}$ The
transition probability $L_h^V$ can be obtained by an analogous
reasoning.

In the case of the Moran process, instead, the probability that in a
timestep the state of node $i$ is flipped from $1$ to $0$ is
\begin{equation}
  P_i (1\rightarrow 0) = \sum_j \frac{s_i\,b_i(1-s_j)}{\av{b}(N-1)}a_j\delta
  t = s_i \,\dfrac{b_i}{\av{b}}\delta  t(\av{a}-\av{a\rho_h}).
\end{equation}
Indeed, the probability that the agent $i$ is flipped from $1$ to $0$
while interacting with $j$ is equal to the probability
$a_j \delta t(1-s_j)$ that $j$ becomes active and is in state $0$, times
the probability $\frac{s_i\,b_i}{\av{b}(N-1)}$ that $i$ is chosen among
all the other agents and is in state $1$. We then sum over all the
agents $j$ to obtain the total probability.  Then, summing over all nodes
$i$ with activity $a$ and attractiveness $b$ we get
\begin{equation}
  L_h^{M} = N_h\,\delta t\,\rho_h\,(\av{a}-\av{a\,\rho_h})\dfrac{b}{\av{b}} \ .
\end{equation}
We obtain in a similar fashion
\begin{equation}
  R_h^{M} = N_h\,\delta t\,\av{a\,\rho_h}\,(1-\rho_h)\dfrac{b}{\av{b}} .
\end{equation}
From these two particular cases we deduce the time evolution equation of
the fraction of nodes with activity $a$ and attractiveness $b$ in state
$1$ in the general mixed case, which is given by
\begin{equation}
  \frac{\partial\rho_h(t)}{\partial
    t}=pa\left(\dfrac{\av{b\,\rho_h}}{\av{b}}-\rho_h\right)+
  (1-p)\av{a}\dfrac{b}{\av{b}}
  \left(\dfrac{\av{a\,\rho_h}}{\av{a}}-\rho_h\right).\label{eq:a}     
\end{equation}

\subsection{Conservation law}

In the case of the voter model on a complete static graph, the total
fraction $\rho$ of voters in state $1$ is conserved by the dynamics.  In
our model, it is clear from the previous equation that this in
not in general true. Nevertheless, we may look for a conserved quantity
of the form
\begin{equation}
\Omega = \sum_h \lambda_h\,\rho_h ,
\label{eq:omega}
\end{equation}
where the weights  $\lambda_h$ are normalized as
$\sum_h \lambda_h = 1$. Using Eq.~\eqref{eq:a}, we can check that the
condition $\partial \Omega / \partial t = 0$ is fulfilled if the
functions $\lambda_h$ satisfy the self consistent equation
\begin{equation}
  \lambda_h =  \eta(h)\,\dfrac{p\,b\,[\sum_{h'} a'\lambda_{h'}]+
    (1-p)\,a\,[\sum_{h'} b'\lambda_{h'}]}{pa\av{b}+(1-p)\av{a}b} ,
\label{eq:lambda_h}
\end{equation}
where $\sum_{h} a\lambda_{h}$ and $\sum_{h} b\lambda_{h}$ are determined
by the normalization of the weights $\lambda_h$ (see details in Appendix):
\begin{equation}
  \sum_{h} a\lambda_{h} = \dfrac{1}{Q_p}\av{\dfrac{a^2}{\Delta_{h,p}}},
  \label{eq:a_lambda}
\end{equation}
\begin{equation}
  \sum_{h} b\lambda_{h}=\dfrac{\av{a}}{Q_p\,\av{b}}\av{\dfrac{b^2}{\Delta_{h,p}}},
\label{eq:b_lambda}
\end{equation}
where we have defined
\begin{equation}
  \Delta_{h,p} = pa\av{b}+(1-p)\av{a}b
  \label{eq:3}
\end{equation}
and 
\begin{equation}
  Q_p=p\av{\dfrac{a^2}{\Delta_{h,p}}}\av{\dfrac{b}{\Delta_{h,p}}}+(1-p)\dfrac{\av{a}}{\av{b}}\av{\dfrac{b^2}{\Delta_{h,p}}}\av{\dfrac{a}{\Delta_{h,p}}}.
  \label{eq:4}
\end{equation}
Notice that this last quantity depends only on $p$.

\subsection{Exit probability}

As in the case of the standard voter model \cite{KineticViewRedner},
the presence of a conservation law allows us to estimate directly the exit
probability $E$ for a single agent with state $1$ in a population
of agents with state $0$, i.e., the probability that all agents finally adopt the state $1$. 
Indeed, the final state with all voters in state $1$,
corresponding to $ \Omega = 1$, takes place with probability $E$ (by definition), while
the final state with all voters in state $0$, with $\Omega = 0$, happens
with probability $1-E$. The conservation of $\Omega$ implies that
$\Omega(t=0) = E \times 1 + (1-E) \times 0$, from where we immediately
obtain
\begin{equation}
  E =\sum_h \lambda_h \,\rho_h(0) ,
  \label{eq:5}
\end{equation}
which depends exclusively on the initial state in which the system is
prepared. For the particular initial conditions consisting of a single
voter with variables $h=(a,b)$, i.e., activity $a$ and attractiveness $b$, in state $1$ in a
background of voters in state $0$, we have that
$\rho_{h'}(0) = \delta_{h', h}\, N_h^{-1}$, which leads to an exit
probability
\begin{equation}
E_{a,b} = \dfrac{\lambda(a,b)}{N\eta(a,b)}, \label{eq:exit}
\end{equation}
which, using Eq.~\eqref{eq:lambda_h} can be more explicitly expressed as
\begin{equation}
  \label{eq:6}
  E_{a,b} = \frac{1}{N Q_p} \frac{p b \av{\dfrac{a^2}{\Delta_{h,p}}} + (1-p)
    a \dfrac{\av{a}}{\av{b}} \av{\dfrac{b^2}{\Delta_{h,p}}}}{p a \av{b} 
    + (1-p) b \av{a}}.
\end{equation}
Interestingly, this exit probability is a function of the ratio
$\dfrac{a}{b}$ only.

\subsection{Average consensus time}

In order to compute the consensus time we can follow
\cite{PhysRevLett.94.178701,Sood08} and apply a one-step calculation to
write down the recursion relation for the time $T[\{\rho_h\}]$ to reach consensus starting
from a configuration $\{\rho_h\}$:
\begin{widetext}
  \begin{equation}
    T[\{\rho_h\}]=\delta t + \left(1-\sum_h
      (R_h+L_h)\right)T[\{\rho_h\}] + \sum_h \left(R_h T[\{\rho_{h'},\rho_h+1/N_h\}]+L_h
      T[\{\rho_{h'},\rho_h-1/N_h\}]\right) ,
    \label{eq:7}
  \end{equation}
\end{widetext}
where the notation $\{\rho_{h'},\rho_h \pm 1/N_h\}$ denotes a
modification of the configuration $\{\rho_h\}$ by the flip of one agent
of variables $h$ (either from state $0$ to the state $1$, for the $+$
case, or vice-versa for the $-$ case).

This equation essentially amounts to consider that the consensus time
for a given configuration is equal to the consensus time at the
configuration obtained after a transition taking place in a time
$\delta t$, weighted by the corresponding transition probabilities, plus
$\delta t$. Expanding Eq.~\eqref{eq:7} at second order in $1/N_h$ we
obtain the backward Kolmogorov equation~\cite{gardiner4ed2010}
\begin{equation}
  \sum_h v_h\frac{\partial T}{\partial \rho_h} + \sum_h
  D_h\frac{\partial^2 T}{\partial \rho_h^2}=-1,
  \label{eq:kolmogorov}
\end{equation}
where 
\begin{equation}
  v_h =pa\left(\dfrac{\av{b\,\rho_h}}{\av{b}}-\rho_h\right) +
  (1-p)\av{a}\dfrac{b}{\av{b}}\left(\dfrac{\av{a\,\rho_h}}{\av{a}}-\rho_h\right) 
\end{equation}
and
\begin{eqnarray}
  D_h &=& \dfrac{pa}{2N_h}\left(\dfrac{\av{b\,\rho_h}}{\av{b}}  +
          \rho_h-2\,\dfrac{\av{b\,\rho_h}}{\av{b}}
          \rho_h\right)\nonumber\\ 
      &+&
          \dfrac{(1-p)\av{a}b}{2\av{b}N_h}\left(\dfrac{\av{a\,\rho_h}}{\av{a}}
          + \rho_h-2\,\dfrac{\av{a\,\rho_h}}{\av{a}} \rho_h\right) 
          \label{eq:drift}
\end{eqnarray}
are the drift and diffusion coefficients,
respectively~\cite{gardiner4ed2010}.  After a transient time depending
on the distribution $\eta(a,b)$, the system reaches a steady state where
$\rho_h=\Omega$, $\forall h$. Then we may drop the drift term in
Eq.~\eqref{eq:kolmogorov}, and, considering Eq.~\eqref{eq:omega}, change variable
from $\rho_h$ to $\Omega$~\cite{PhysRevLett.94.178701,Sood08}
\begin{equation}
  \frac{\partial T}{\partial \rho_h} = \lambda_h\,\frac{\partial T}{\partial \Omega} .
  \label{eq:change_variable}
\end{equation}
Substituting into Eq.~\eqref{eq:kolmogorov} and simplifying (see details
in Appendix), we finally obtain
\begin{equation}
 \Omega(1-\Omega)\,\frac{\partial^2
    T}{\partial \Omega^{2} }=\dfrac{-N\av{b}}{\left(\sum_h a\lambda_h\right)\left(\sum_h b\lambda_h\right)} .
    \label{eq:evolT}
\end{equation}
This last equation can be directly integrated, yielding
the consensus time
\begin{equation}
  T=\tau\dfrac{N}{\av{a}}\left((1-\Omega)\ln\frac{1}{1-\Omega}+\Omega\ln\frac{1}{\Omega}\right) ,
\end{equation}
where we defined the characteristic adimensional consensus time
\begin{equation}
\tau =\dfrac{\av{a}\av{b}}{\left(\sum_h a\lambda_h\right)\left(\sum_h b\lambda_h\right)}.
\label{eq:tau}
\end{equation}
The model is then entirely solved in terms of the previous expressions for
the consensus time and the exit probability. These expressions are however
 quite intricate and it is quite insightful to study particular cases of
interest, for given forms of the distribution of the activity and
attractiveness $\eta(a,b)$ and particular values of the mixing
probability $p$. We present this analysis in the following
Section.

\section{Particular cases}
\label{sec:part.cases}

\subsection{p = 1/2}

From the definition of  $\Delta_{h,p}$ in
Eq.~\eqref{eq:3}, one obtains, by multiplying this equation respectively by $a/\Delta_{h,p}$ and
$b/\Delta_{h,p}$ and averaging,
\begin{eqnarray}
\av{b}&=& 
p \av{\dfrac{ab}{\Delta_{h,p}}}\av{b}+ (1-p)\av{\dfrac{b^2}{\Delta_{h,p}}}\av{a}\label{eq:avg_b}, \\
\av{a}&=&  
p \av{\dfrac{a^2}{\Delta_{h,p}}}\av{b}+(1-p)\av{\dfrac{ab}{\Delta_{h,p}}}\av{a}.
\end{eqnarray}
Thus for $p=1/2$, one obtains, eliminating $\langle ab/\Delta_{h,p} \rangle$ between these two equations,
\begin{equation}
\dfrac{\av{a}}{\av{b}}\av{\dfrac{b^2}{\Delta_{h,p}}} =
  \dfrac{\av{b}}{\av{a}}\av{\dfrac{a^2}{\Delta_{h,p}}} ,
\end{equation}
and Eq.~\eqref{eq:4} becomes
\begin{eqnarray}
Q_{1/2}&=& \dfrac{1}{2}\av{\dfrac{b^2}{\Delta_{h,p}}}\dfrac{\av{a}}{\av{b}}\left(\av{\dfrac{a}{\Delta_{h,p}}}+\av{\dfrac{b}{\Delta_{h,p}}}\dfrac{\av{a}}{\av{b}}\right)\nonumber\\
&=& \av{\dfrac{b^2}{\Delta_{h,p}}}\dfrac{\av{a}}{\av{b}^2}=\dfrac{\av{1}}{\av{a}}\av{\dfrac{a^2}{\Delta_{h,p}}}.
\end{eqnarray}
From here, it follows that $\sum_h a\,\lambda_h=\av{a}$ and $\sum_h b\,\lambda_h=\av{b}$, which finally implies that $\lambda_h=\eta_h$ and $\tau = 1$. 

In this case, the dynamics becomes identical to the standard link update
dynamics of the voter model~\cite{Suchecki05}, and it is totally
independent on $a$ and $b$ (in terms of the number of flip attempts)
because the probability that the total number of voters in state $1$ is
increased during an update attempt is exactly compensated by the
probability that this same number is decreased. 

\subsection{Pure voter model}

The voter model, corresponding to $p=1$, leads in Eq.~\eqref{eq:lambda_h}
to
\begin{equation}
  \lambda(a,b) = \eta(a,b) \dfrac{b\,a^{-1}}{\av{b\,a^{-1}}} \ .
\end{equation}
We also obtain $\Delta_{h,p} = a \av{b}$ and $Q_1 = \av{a} \av{b a^{-1}} / \av{b}^2$,
leading for the consensus time in Eq.~\eqref{eq:tau} to the simple form:
\begin{equation}
\tau = \av{a}\dfrac{\av{b\,a^{-1}}^2}{\av{b^2 a^{-1}}}
\label{eq:voter_tau}
\end{equation}
The exit probability is also straightforward to derive from Eq.~\eqref{eq:exit}:
\begin{equation}
E_{a,b} = \dfrac{b\,a^{-1}}{N\av{b\,a^{-1}}} \label{eq:voter_exit} .
\end{equation}

\subsection{Moran process}

The Moran process corresponds to $p=0$, then Eq.~\eqref{eq:lambda_h} reduces to
\begin{equation}
\lambda(a,b) = \eta(a,b) \dfrac{a\,b^{-1}}{\av{a\,b^{-1}}},
\end{equation}
leading, with $\Delta_{h,p}= b \av{a}$ and $Q_0 = \langle a b^{-1} \rangle/\av{a}$, from Eq.~\eqref{eq:tau} to
\begin{equation}
\tau = \av{b}\dfrac{\av{a\,b^{-1}}^2}{\av{a^2 b^{-1}}}.
\label{eq:moran_tau}
\end{equation}
The exit probability reads in this case
\begin{equation}
E_{a,b} = \dfrac{a\,b^{-1}}{N\av{a\,b^{-1}}}. \label{eq:moran_exit}
\end{equation}
It is noteworthy that the results for the Moran process are obtained from the ones of the voter
model by simply exchanging $a$ and $b$. In fact, we see from
Eq.~\eqref{eq:lambda_h} that the dynamics of the mixed process is the same
as the dynamics of the symmetrical process (i.e. with $p\leftarrow 1-p$)
upon exchanging $a$ and $b$. This is intuitively clear if we examine the process
from a stochastic point of view: at each update attempt, the node $i$ is
chosen at random with probability $\frac{a_i}{\av{a}}$ and the node $j$
with probability $\frac{b_j}{\av{b}}$. Besides, changing $p$ into $1-p$
is equivalent to reversing the roles of $i$ and $j$, which has no effect
if $a$ and $b$ are exchanged. This is however valid only when the time
is measured as the number of update attempts, the physical time being
multiplied by $\frac{\av{a}}{\av{b}}$ when swapping $a$ and $b$.

\subsection{Pure Activity Driven Networks}

The original activity driven network model~\cite{2012arXiv1203.5351P}
does not consider a heterogeneous attractiveness, and this corresponds
to a joint distribution $\eta(a,b) = F(a) \delta_{b, b_0}$, where $F(a)$
is the activity distribution and $b=b_0$, constant. In this case we have
$Q_p=\dfrac{\av{a}}{b_0}\av{\,[pa+(1-p)\av{a}]^{-1}}$, and the characteristic consensus time reads
\begin{equation}
\label{eq:taugeneral}
\tau =
\dfrac{\av{a}^2\av{\,[pa+(1-p)\av{a}]^{-1}}}{\av{a^2\,[pa+(1-p)\av{a}]^{-1}}} ,
\end{equation}
while the exit probability is given by
\begin{equation}
E_a = \dfrac{1}{N}\dfrac{p\frac{\av{a}}{\tau}+(1-p)a}{pa+(1-p)\av{a}} .
\end{equation}

\begin{figure}[t]
  \centering
  \includegraphics[width=8cm]{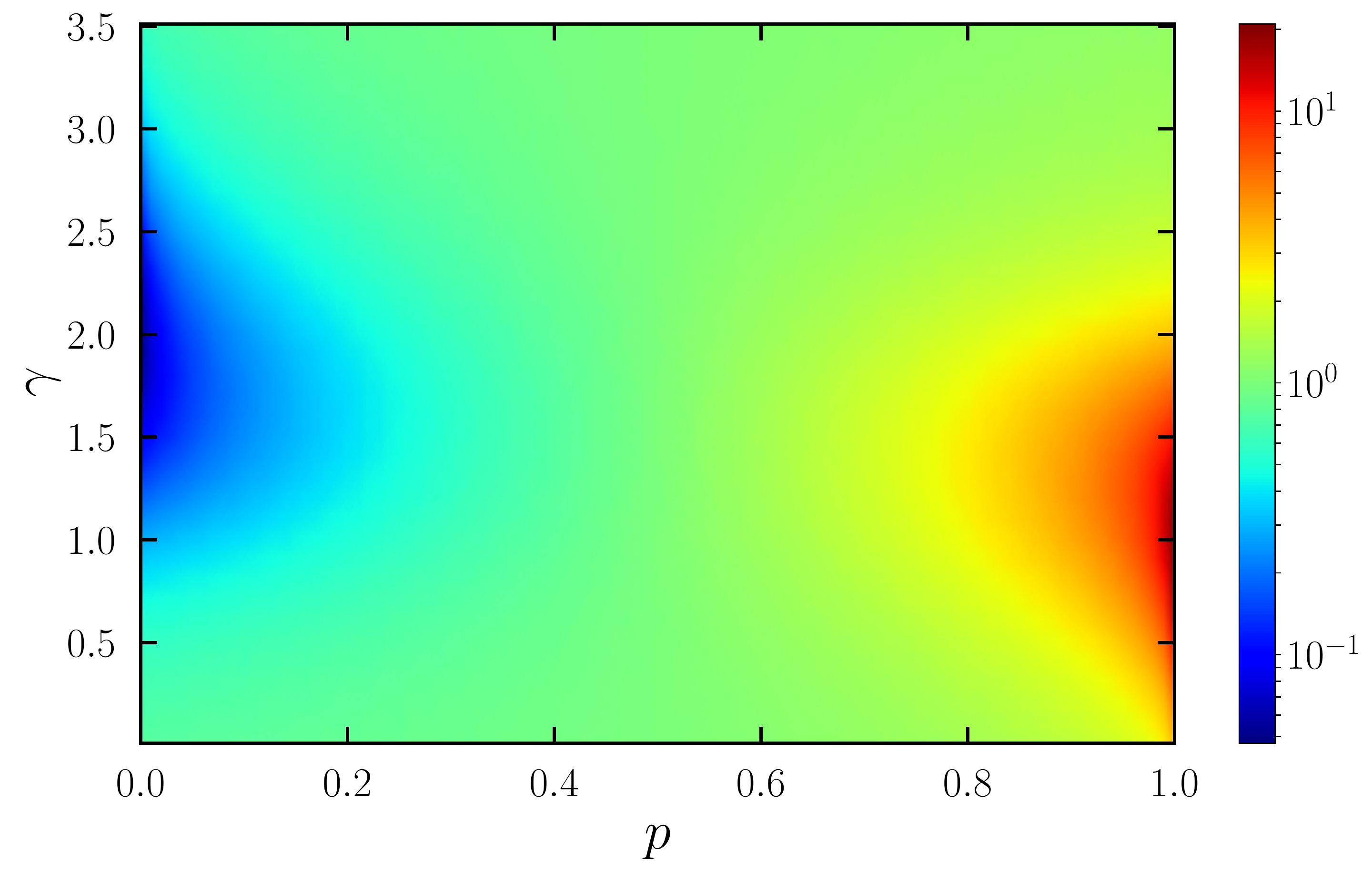}
  \caption{Characteristic consensus time $\tau$ for the dynamics on a pure activity driven network, i.e., fixed attractivity $b=b_0$,
    and a distribution of activities $F(a)$ given by Eq. \eqref{eq:eta_distr}, as a
    function of $p$ and of the exponent $\gamma$ of the distribution $F$. $\epsilon = 10^{-3}$.}
  \label{fig:bequal1}
\end{figure}

In order to study the behavior of the consensus time, in
Fig.~\ref{fig:bequal1} we plot the analytical evaluation of $\tau$, Eq. \eqref{eq:taugeneral}, for a
normalized activity distribution with a power-law form, as empirically
observed in Ref. \cite{2012arXiv1203.5351P},
\begin{equation}
  F(a) = \frac{1-\gamma}{1-\epsilon^{1-\gamma}} a^{-\gamma}, \quad a
  \in [\epsilon, 1].
  \label{eq:eta_distr}
\end{equation}
where $\epsilon$ is the minimum activity in the system, imposed in order
to avoid divergences in the normalization and moments of $F(a)$. From
Fig.~\ref{fig:bequal1} we see that the consensus time has a minimum
around $\gamma=2$ for the Moran process ($p=0$) and a maximum around
$\gamma=1$ for the voter model ($p=1$). Note that by virtue of the
symmetry property discussed above, the dynamics of the pure
attractiveness model (setting $a_i=a_0, \forall i$), taking the same
distribution $F$ for $b$ and imposing $a_0=b_0$, is the same upon
exchanging $p$ by $1-p$. In particular, the consensus time is obtained
by reversing the $p$ axis in Fig.~\ref{fig:bequal1}.

\subsection{Independent activity and attractiveness}

\begin{figure}[t]
  \centering
  \includegraphics[width=8cm]{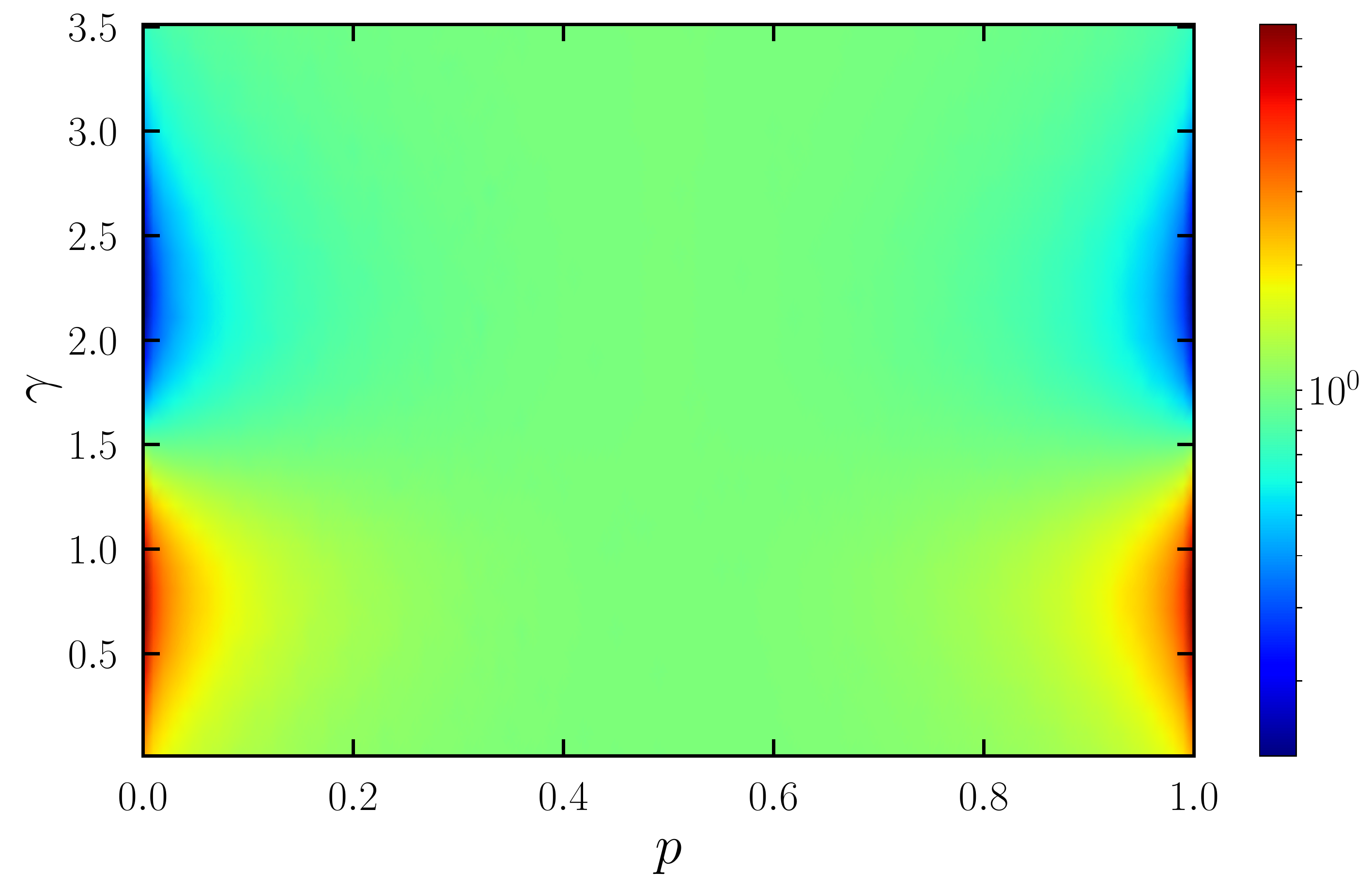}
  \caption{Characteristic consensus time $\tau$ as a function of
    $\gamma$ and $p$ in the case $\eta(a,b) = F(a) F(b)$, with $F$ given by Eq. \eqref{eq:eta_distr} and $\epsilon = 10^{-3}$.}
  \label{fig:bindpdt}
\end{figure}

In the case where $a$ and $b$ are drawn independently from the same
distribution $F$, we have $\eta(a,b) = F(a) F(b)$. 
In Fig.~\ref{fig:bindpdt} we plot the characteristic
consensus time $\tau$ as a function of $p$ and $\gamma$ for $F$ given by Eq.~\eqref{eq:eta_distr}.
For this particular form of the distribution $\eta(a,b)$ (and in general
for any symmetric joint distribution such that $\eta(a,b)=\eta(b,a)$),
the dynamics remains the same when changing $p$ into $1-p$ because
exchanging $a$ and $b$ has no effect. This is clearly observed in
Fig.~\ref{fig:bindpdt}. Additionally, we see that both the voter model
and the Moran process have a minimum consensus time for
$\gamma \simeq 2.25$ and a maximum consensus time for $\gamma = 0.75$,
respectively.

\subsection{Strongly correlated activity and attractiveness}

As we previously mentioned, the weight function $\lambda_h$ is the
product of $\eta_h$ and a function of the ratio $\frac{a}{b}$. This fact
straightforwardly implies that, in the maximally correlated case
$\eta(a,b) = F(a) \delta_{a,b}$, where $a=b$ for every agent, the
dynamics is the same as in a fully connected static network, i.e. the
average density $\rho=\av{\rho_h}$ of voters in state $1$ is conserved,
the reduced consensus time $\tau$ is equal to $1$, and the exit
probability is homogeneous and equals $1/N$.

\subsection{Discussion}

The results obtained above relate the average consensus time and the
exit probability with the moments of the joint distribution $\eta(a,b)$
and the value of the activity $a$ and the attractiveness $b$ of the
initial invading voter. 
Remarkably, when we compare these results with
the ones obtained in the case of static networks with a given degree
distribution $P(k)$~\cite{Sood08,PhysRevLett.94.178701} we observe
interesting symmetries between voter and Moran dynamics. 

This symmetry is of particular interest when we consider the pure activity driven network (setting $b=1$). Let us consider
for instance the invasion exit probability,
Eqs.~\eqref{eq:voter_exit} and~\eqref{eq:moran_exit}. In the case of the voter
model, this exit
probability is inversely proportional to the activity of the node with initial state $1$ ($E^{voter}_a \propto 1/a$), while for the
Moran process, it is proportional to the activity ($E^{Moran}_a \propto a$). 
This can be understood by the fact that an active node will often change state in the voter model, by contacting other nodes,
while in the Moran process it will often spread his/her state towards the other nodes contacted.

These results have to be compared with the result for the voter and Moran processes in static
networks, in which the exit probability for a single node of degree $k$ with state $1$ 
is $E^\mathrm{voter}_k \sim k $ for the voter and
$E^\mathrm{Moran}_k \sim k^{-1}$ for the Moran process~\cite{Sood08,PhysRevLett.94.178701}. 
Intuitively indeed, in the case of static networks, in the voter model high degree nodes are
chosen to be copied with high probability~\cite{Dorogovtsev:2002}, implying that they are very efficient spreaders of their own state to the
rest of the network. Hence, the larger $k$, the higher the exit probability. In the case of the Moran process, by applying the
same argument, high degree nodes are prone to often change state by adopting the state of a neighbor \cite{Sood08,PhysRevLett.94.178701}, 
and hence the exit probability decreases with the degree.

Let us now recall that, for a pure activity
driven network, the aggregated degree of a node with activity $a$ takes the value
$\bar{k}_a(t) \sim (a+\av{a}) t$ at time $t$: nodes with high activity tend
to have large integrated degree~\cite{2012arXiv1203.5351P,starnini_topological_2013}. 
Putting this in relation with the behavior of the exit probability as a function of activity in temporal networks and of degree in 
static networks, we thus obtain that the dynamics on the temporal activity driven network 
  yields a completely
different and opposite result when compared with the dynamics on the static, integrated network counterpart:
High activity nodes are more prone to spread under Moran dynamics, while
low activity nodes are more prone under voter dynamics. 

This symmetry
voter-Moran between pure activity driven networks and their integrated
counterpart occurs as well at the level of the  average consensus time when we
measure it as a function of the update attempts. Considering that a
randomly chosen node becomes active with average probability $\av{a}$,
we have that, as a function of updated attempts, the convergence time is
$\bar{T}_N \equiv \av{a} T_N$. We have thus, for homogeneous initial
conditions ($\Omega=1/2$),
\begin{equation}
  \bar{T}^\mathrm{voter}_N = N \av{a} \av{a^{-1}}\ln 2, \quad 
  \bar{T}^\mathrm{Moran}_N =  N \frac{\av{a}^2} {\av{a^{2}}}\ln 2.
  \label{eq:13}
\end{equation}
Comparing with the results for static
networks~\cite{Sood08,PhysRevLett.94.178701},
\begin{equation}
  \bar{T}^\mathrm{Moran}_N = N \av{k} \av{k^{-1}}\ln 2, \quad 
  \bar{T}^\mathrm{voter}_N =  N \frac{\av{k}^2} {\av{k^{2}}}\ln 2,
  \label{eq:14}
\end{equation}
we observe that the formulas for voter and Moran
dynamics are indeed mirror images, with the activity distribution $a$ in the
temporal representation substituted by the degree distribution in the
integrated representation. 

Let us now consider instead the pure attractiveness temporal network model (setting $a=1$).
 In that case,  the exit probability is proportional to the attractiveness for the voter model, $E_b^{voter} \propto b$,
while for the Moran process, it is inversely proportional to the
attractiveness, $E_b^{Moran} \propto 1/b$. Moreover, the integrated degree of a node
with attractiveness $b$ is $\bar{k}_b(t) \sim \frac{b}{\av{b}}t$. 
Here therefore, we have the same kind of behavior on the temporal and corresponding integrated static network
when making an equivalence between attractiveness in the temporal network and degree in the static network.
This equivalence between a static network with a degree distribution $P(k)$
and a pure attractiveness temporal network with the same distribution
$P(b)$ is also obtained by looking at the consensus time measured as the number
of update attempts.

\section{Numerical results}
\label{sec:numerics}

In order to check the analytical predictions made above, we have
performed simulations of the mixed process defined earlier on activity
driven networks with attractiveness, choosing a marginal activity
distribution following a power-law, Eq.~\eqref{eq:eta_distr}, similar to
the distribution observed
empirically in some real networks \cite{2012arXiv1203.5351P}.
We have performed simulations for network sizes $N=10^2$, $10^3$ and
$10^4$, averaging over $10^3$ realizations. 

\begin{figure}[t]
  \centering
  \includegraphics[width=8cm]{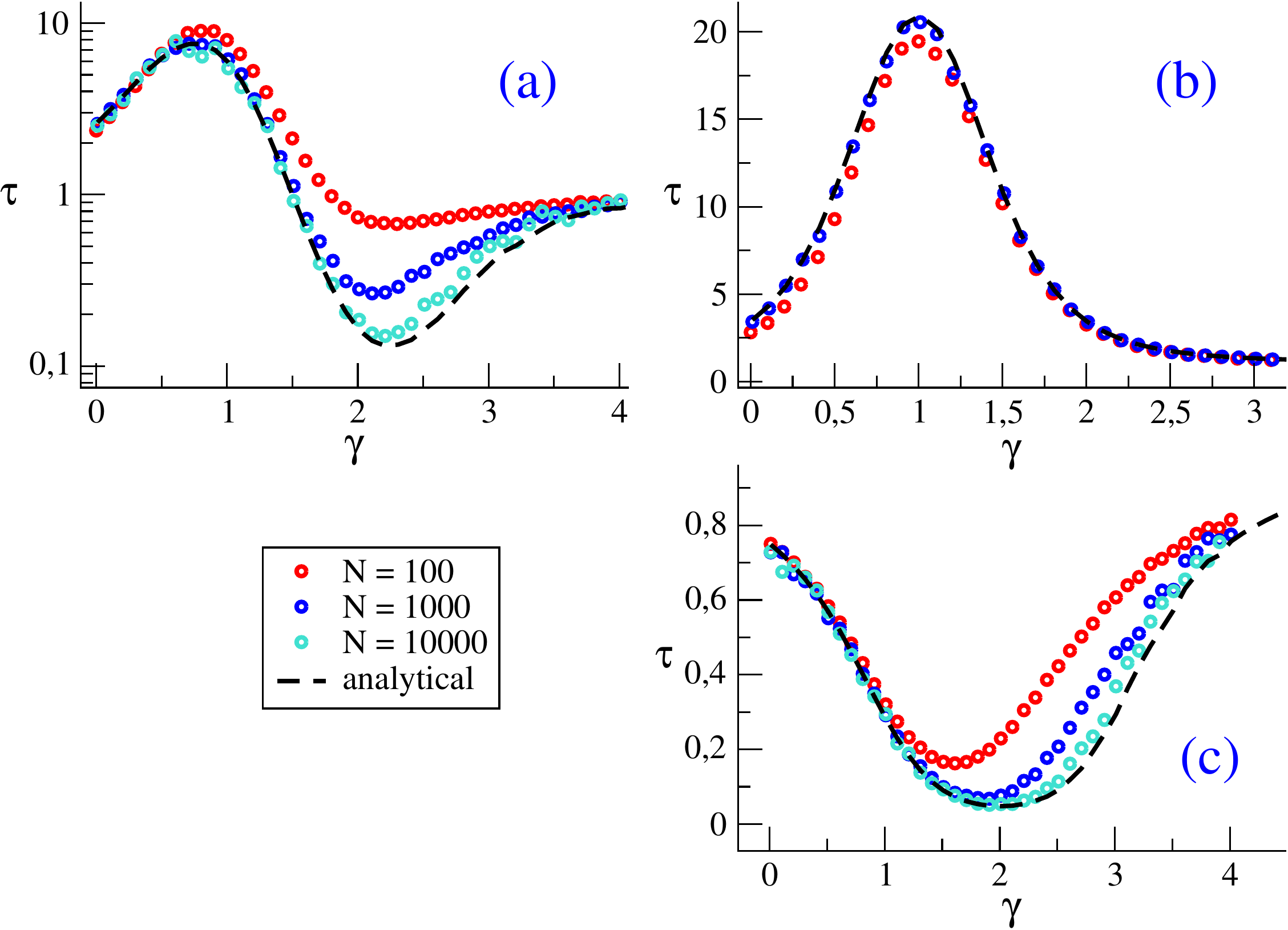}
  \caption{Voter-like dynamics in temporal activity driven networks with attractiveness. Reduced consensus time as
    a function of $\gamma$ for different values of the network size
    $N$, for an activity distribution $F(a)$ given by Eq.~\eqref{eq:eta_distr} with $\epsilon = 10^{-3}$.    
     (a): Voter and Moran processes (equivalent) for $\eta(a,b)=F(a)F(b)$. (b): Voter model on pure AD network ($b=1$). (c): Moran process on pure AD network ($b=1$).
     In each case, the dashed line corresponds to the analytical expression given by Eq.~\eqref{eq:tau}. 
     }
  \label{fig:size_scaling}
\end{figure}

In Fig.~\ref{fig:size_scaling} we plot the reduced consensus time
$\tau$ as a function of $\gamma$ for three different values of the
network size and three different  dynamics: voter and
Moran processes on a pure activity driven network, and voter model on an
activity driven network with attractiveness with $a$ and $b$
independently and equally distributed $\eta(a,b)=F(a)F(b)$. The curves
are compared to the theoretical value given in Eq.~\eqref{eq:tau}. We see
that for $N=10^4$ the dynamics already matches well the expected behaviour in
the infinite size limit. We deduce that our heterogeneous mean-field analysis 
captures efficiently the opinion dynamics on the activity driven network with attractiveness.

\section{Asymptotic behaviour}
\label{sec:asympt}

In Figs.~\ref{fig:bequal1}-\ref{fig:size_scaling}, we see that the
consensus time presents minima and maxima when $\gamma$ varies.  To
investigate this point in more details, we analyse the asymptotic
behaviour of the moments of the distribution $F(a)$ when $\epsilon$
tends to zero. For an activity distributed with
Eq.~\eqref{eq:eta_distr}, the moments of $a$ take the form
\begin{equation}
  \av{a^n} = \frac{1-\gamma}{n+1-\gamma}
  \frac{1-\epsilon^{1+n-\gamma}}{1-\epsilon^{1-\gamma}}.
  \label{eq:moments}
\end{equation}
The dynamics of the voter and Moran processes on pure activity driven
network and on activity driven network with equally and independently
distributed $a$ and $b$ depends on the moments $\av{a^{-1}}$, $\av{a}$
and $\av{a^2}$ only. The asymptotic behaviour of these three quantities for $\epsilon \to 0$
are summarized in Tables~\ref{tab:asymptotic} and \ref{tab:2}, along 
with the resulting behaviour of the consensus time in the case with $a$ and $b$ independently
and equally distributed and $p=1$ (or equivalently $p=0$), for which 
\begin{equation}
\tau = \dfrac{\av{a}^3\av{a^-1}}{\av{a^2}}.
\label{eq:tau_ab_indpdt_voter}
\end{equation}

\begin{figure}[t]
  \centering
  \includegraphics[width=8cm]{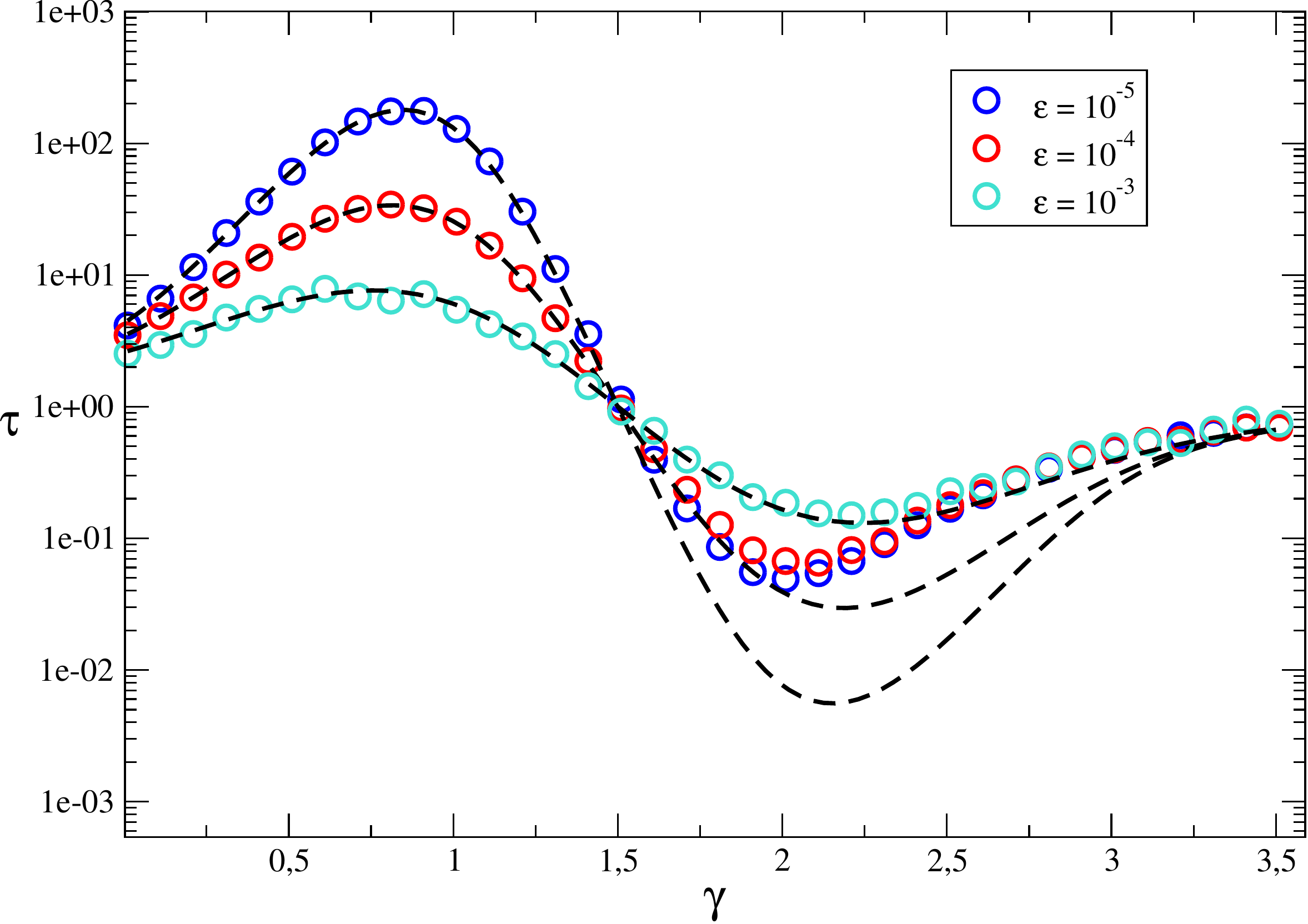}
  \caption{Voter dynamics in temporal networks with independent and equally distributed
    activity and attractiveness. Consensus time $\tau$ as a function of
    $\gamma$ for activity and attractiveness distributed according to Eq.~\eqref{eq:eta_distr}
    and different values of $\epsilon$. The analytical expression
    given by Eq.~\eqref{eq:tau_ab_indpdt_voter} is shown in dashed lines. The numerical simulations are performed with network size
    $N = 10^{4}$.}
  \label{fig:epsilon}
\end{figure}

 We observe that for $0<\gamma<1.5$ the consensus time in Eq.~\eqref{eq:tau_ab_indpdt_voter}
diverges when $\epsilon$ goes to zero, and tends instead to zero for
$1.5<\gamma<3$. We also recover the fact that the fastest consensus is
reached for $\gamma = 2$ and the slowest for $\gamma = 1$ and that for
$0<\gamma<3$ the consensus time exhibits a symmetry with respect to the
axis $\gamma=1.5$ : $\tau(\gamma)=\frac{1}{\tau(3-\gamma)}$. Finally,
for  $\gamma \gg 3$, the
heterogeneity of the distribution of $a$ is no longer significant, so
that the dynamics is that of a fully connected static network.  In
Fig.~\ref{fig:epsilon} we plot the reduced consensus time for the voter
dynamics on a network with independent activity and attractiveness as obtained by direct numerical simulations
of a voter model on a temporal network, compared with the analytical predictions of Eq.~\eqref{eq:tau_ab_indpdt_voter}, for various values of $\epsilon$ and $N=10^4$. The
simulations confirm the predicted asymptotic behaviour of $\tau$ when $\epsilon$
tends to zero. We also observe stronger finite size effects when epsilon
tends to zero due to a poorer sampling of the activity distribution given by Eq.~\eqref{eq:eta_distr}.

\begin{table}[t]
  \begin{ruledtabular} 
    \begin{tabular}{|c||c|c|c|c|}
   $\gamma$ & $]0,1[$  & $]1,2[$   & $]2,3[$ & $>3$\\ \hline  
      $\av{a^{-1}}$   & ${\mathcal O}(\epsilon^{-\gamma})$   & ${\mathcal O}(\epsilon^{-1})$  & ${\mathcal O}(\epsilon^{-1})$& ${\mathcal O}(\epsilon^{-1})$\\
\hline  
      $\av{a}$  &  ${\mathcal O}(1)$  & ${\mathcal O}(\epsilon^{\gamma-1})$ & ${\mathcal O}(\epsilon)$& ${\mathcal O}(\epsilon)$\\ \hline  
      $\av{a^{2}}$   & ${\mathcal O}(1)$ &${\mathcal O}(\epsilon^{\gamma-1})$ &${\mathcal O}(\epsilon^{\gamma-1})$& ${\mathcal O}(\epsilon^2)$\\ \hline  \hline
      $\tau$  & ${\mathcal O}(\epsilon^{-\gamma})$   & ${\mathcal O}(\epsilon^{2\gamma-3})$  & ${\mathcal O}(\epsilon^{3-\gamma})$& ${\mathcal O}(1)$ \\     
    \end{tabular}
  \end{ruledtabular} 
  \caption{Asymptotic behaviour of the moments of the distribution $F(a)$ defined in Eq.~\eqref{eq:eta_distr}, and resulting
  asymptotic behaviour of the reduced consensus time $\tau$ given by Eq.~\eqref{eq:tau_ab_indpdt_voter} when $\epsilon$ tends to zero, as a function of the exponent $\gamma$.
  For $\gamma=0, 1, 2, 3$, logarithmic corrections are present, given in Table \ref{tab:2}.
  }
  \label{tab:asymptotic}
\end{table}

\begin{table}[t]
  \begin{ruledtabular} 
    \begin{tabular}{|c||c|c|c|c|}
   $\gamma$ & 0 &  1 &  2 &  3 \\ \hline  
      $\av{a^{-1}}$  & $-\ln \epsilon$ &  $-(\epsilon\ln\epsilon)^{-1}$  & $(2\epsilon)^{-1}$ & $\frac{2}{3}\epsilon^{-1}$\\
\hline  
      $\av{a}$  &  $1/2$   & $-(\ln\epsilon)^{-1}$& $-\epsilon\ln\epsilon$ & $2\epsilon$\\ \hline  
      $\av{a^{2}}$  &  1/3   & $-(2\ln\epsilon)^{-1}$& $\epsilon$ & $-2\epsilon^2\ln\epsilon$ \\ \hline  \hline
      $\tau$  & $-\frac{3}{8}\ln\epsilon$ & $-2\epsilon^{-1}(\ln\epsilon)^{-3}$ &  $-\frac{1}{2}\epsilon(\ln\epsilon)^3$  &  $-\frac{8}{3}(\ln\epsilon)^{-1}$\\     
    \end{tabular}
  \end{ruledtabular} 
  \caption{Asymptotic behaviour of the moments of the distribution $F(a)$ defined in Eq.~\eqref{eq:eta_distr}, and resulting
  asymptotic behaviour of the reduced consensus time $\tau$ given by Eq.~\eqref{eq:tau_ab_indpdt_voter} when $\epsilon$ tends to zero, for the specific cases $\gamma = 0, 1, 2, 3$.}
  \label{tab:2}
\end{table}

\section{Conclusions}

In this paper, we have studied in detail the properties of consensus
processes mixing the voter and Moran models update rules, on temporal
network models based on the activity driven paradigm.  Through a
heterogeneous mean-field approach, we have derived the evolution
equation of the average density of voters in state $1$ with activity $a$
and attractiveness $b$. This has allowed us to identify a conserved
quantity of the dynamics, and subsequently to compute the average time
to reach consensus and the probability that a single agent with a
discrepant opinion among an otherwise unanimous population spreads
his/her opinion to the whole network, called the exit
probability. Surprising results arise from the study of particular cases
of the distribution of the parameters $a$ and $b$. When the
attractiveness is taken to be proportional to the activity $a$, the
dynamics is the same as if the copying process were running on a static
complete graph. The average activity $\av{a}$ determines the time scale
of the dynamics, but otherwise the precise distribution of activity
among the agents is no longer relevant. This holds for all values of the
probability $p$ determining the state update rule, and in particular for
the voter model ($p=1$) and the Moran process ($p=0$). The same
behaviour happens when $p=1/2$, regardless of the distribution
$\eta(a,b)$ of activity and attractiveness: surprisingly, the exit
probability is equal to $1/N$ and does not depend of the parameters of
the initial invading node.

Interestingly, when the activity and the attractiveness are independent
and equally distributed, the dynamics is unchanged when replacing $p$ by
$1-p$. In fact, it appears that by construction, when the time is
counted as the number of update attempts, exchanging $a$ and $b$ on all
the nodes is equivalent to replace $p$ by $1-p$ in the update rule. One
of our main results lies in the observation that the voter model and the
Moran process on a pure activity driven network (setting $b=1$ for all
nodes) are in some sense mirror images of their static network
counterparts. Indeed, the dynamics of the voter model on an activity
driven network with a distribution $F(a)$ is the same as the Moran
dynamics running on top of a static network with a degree distribution
$P(k)=F(k)$. The same holds for the Moran process on the activity driven
network and the voter model on the static network. This implies that the
apparently appealing operation consisting in considering an activity
driven network and its integrated counterpart as similar substrates for
this kind of opinion dynamics process would be misleading, despite the
fact that the degree distribution of the integrated network is
practically equal to the activity distribution of the temporal network
\cite{2012arXiv1203.5351P}. On the contrary, a pure attractiveness
temporal network (setting $a=1$ for all nodes) and and its integrated
counterpart are equivalent substrates for the voter and Moran
processes. It would be very interesting to check whether similar
conclusions hold for other consensus formation processes with more
complex update rules like the majority rule process. Our results will
hopefully motivate further research in this direction.
\label{sec:conclusions}

\appendix
\section{Details of some computations}

From the expression of the weights $\lambda_h$ in
Eq.~\eqref{eq:lambda_h} we obtain, by multiplying by $a$ and summing
over $h$ (in the next equations we write $\sum a \lambda$ and
$\sum b \lambda$ as shorthands for $\sum_h a \lambda_h$ and
$\sum_h b \lambda_h$, respectively):
\begin{equation}
\sum_h a\lambda_h = p \,\av{\dfrac{ab}{\Delta_{h,p}}}\,\left[\sum a\lambda\right]+(1-p) \,\av{\dfrac{a^2}{\Delta_{h,p}}}\,\left[\sum b\lambda\right],
\label{eq:sum_a_lambda0}
\end{equation}
which gives a relation between $\sum a\lambda$ and $\sum b\lambda$
\begin{equation}
\sum b\lambda =
\dfrac{1-p\av{\dfrac{ab}{\Delta_{h,p}}}}{(1-p)\av{\dfrac{a^2}{\Delta_{h,p}}}}
\sum a\lambda .
\label{eq:sum_b_lambda}
\end{equation}

The normalization of the weights gives
\begin{equation}
 p \,\av{\dfrac{b}{\Delta_{h,p}}}\,\left[\sum a\lambda\right]+(1-p) \,\av{\dfrac{a}{\Delta_{h,p}}}\,\left[\sum b\lambda\right] = 1 \ .
 \label{eq:sum_lambda}
\end{equation}
Combining Eqs.~\eqref{eq:sum_b_lambda} and \eqref{eq:sum_lambda} leads to
\begin{equation}
\sum a \lambda = \dfrac{\av{\dfrac{a^2}{\Delta_{h,p}}}}{p\av{\dfrac{b}{\Delta_{h,p}}}\av{\dfrac{a^2}{\Delta_{h,p}}}+
\left(1-p\av{\dfrac{ab}{\Delta_{h,p}}}\right)\av{\dfrac{a}{\Delta_{h,p}}}}.
\label{eq:sum_a_lambda}
\end{equation}
Besides, by definition of $\Delta_{h,p}$ we wrote Eq.~\eqref{eq:avg_b}
which, after dividing by $\av{b}$ gives
\begin{equation}
1-p\av{\dfrac{ab}{\Delta_{h,p}}}= (1-p)\av{\dfrac{b^2}{\Delta_{h,p}}}\dfrac{\av{a}}{\av{b}}.
\end{equation}
Inserting this into Eqs.~\eqref{eq:sum_a_lambda} and \eqref{eq:sum_b_lambda}
one recovers the correct expressions given in Eqs.~\eqref{eq:a_lambda} and \eqref{eq:b_lambda}. 

To derive the expression of the average consensus time, we write, combining Eqs.\eqref{eq:kolmogorov}, \eqref{eq:drift} and \eqref{eq:change_variable}
\begin{equation}
 \Omega(1-\Omega)\,\frac{\partial^2
    T}{\partial \Omega^{2} }\sum_h \eta_h\, \Delta_{h,p}\left(\dfrac{\lambda_h}{\eta_h}\right)^2=-N\av{b}.
    \label{eqAppA}
\end{equation}

We have
\begin{widetext}

\begin{eqnarray}
\sum_h \eta_h\, \Delta_{h,p} \left(\dfrac{\lambda_h}{\eta_h}\right)^2 &=& \sum_h \eta_h\,\dfrac{(pb\left[\sum a \lambda\right]+(1-p)a\left[\sum b \lambda\right])^2}{\Delta}\nonumber\\\nonumber\\
&=& \left[\sum a \lambda\right]\;\left[\sum b \lambda\right]\left(p^2\av{\dfrac{b^2}{\Delta_{h,p}}}
\dfrac{\sum a\lambda}{\sum b \lambda}+(1-p)^2\av{\dfrac{a^2}{\Delta_{h,p}}}
\dfrac{\sum b\lambda}{\sum a \lambda}+2p(1-p)\av{\dfrac{ab}{\Delta_{h,p}}}\right)\nonumber\\\nonumber\\
&= & \dfrac{\left[\sum a \lambda\right]\;\left[\sum b \lambda\right]}{\av{a}\av{b}}\left(p^2\av{\dfrac{a^2}{\Delta_{h,p}}}\av{b}^2+(1-p)^2\av{\dfrac{b^2}{\Delta_{h,p}}}\av{a}^2+2p(1-p)\av{\dfrac{ab}{\Delta_{h,p}}}\av{a}\av{b}\right)\nonumber\\\nonumber\\
&=& \dfrac{\left[\sum a \lambda\right]\;\left[\sum b \lambda\right]}{\av{a}\av{b}}\left\langle\dfrac{(\,pa\av{b}+(1-p)\av{a}b\,)^2}{\Delta_{h,p}}\right\rangle\nonumber\\\nonumber\\
&=&  \left[\sum a \lambda\right]\;\left[\sum b \lambda\right] ,
\end{eqnarray}
\end{widetext}
which, inserted in Eq.~\eqref{eqAppA}, finally yields Eq.~\eqref{eq:evolT}.

\begin{acknowledgments}
  We acknowledge financial support from the Spanish MINECO, under
  project FIS2016-76830-C2-1-P.  R.P.-S. acknowledges additional
  financial support from ICREA Academia, funded by the Generalitat de
  Catalunya.
\end{acknowledgments}

\end{document}